\documentclass[aps,twocolumn,prl]{revtex4}
\usepackage{epsfig}
\newcommand{\be}{\begin{equation}}
\newcommand{\ee}{\end{equation}}
\newcommand{\ba}{\begin{array}}
\newcommand{\ea}{\end{array}}
\newcommand{\tq}{\tilde{q}}

\begin{document}
\title{A Growth-based Optimization Algorithm for Lattice Heteropolymers}
\author{Hsiao-Ping Hsu, Vishal Mehra, Walter Nadler, and Peter Grassberger}
\affiliation{John-von-Neumann Institute for Computing, Forschungszentrum
J\"ulich, D-52425 J\"ulich, Germany}

\date{\today}
\begin{abstract}
An improved version of the pruned-enriched-Rosenbluth method (PERM) is
proposed and tested on finding lowest energy states in simple models of
lattice heteropolymers.
It is found to outperform not only the previous version of PERM, but also all
other fully blind general purpose stochastic algorithms which have been
employed on this problem. In many cases it found new lowest energy states
missed in previous papers. Limitations are discussed.
\end{abstract}
\maketitle

Lattice polymers have been studied intensively to understand
protein folding, one of the central problems of computational biology.
A popular model used in these studies is the so-called HP
model \cite{Dill85,Lau89} where only two types of monomers,
H (hydrophobic) and P (polar) ones, are considered.
Hydrophobic monomers
tend to avoid water which they can only by mutually attracting themselves.
The polymer is modeled as a self-avoiding chain on a regular (square or
simple cubic) lattice with interactions
$(\epsilon_{HH},\epsilon_{HP},\epsilon_{PP})=-(1,0,0)$
between neighboring non-bonded monomers.

This model might be too simple to represent finer details of real protein
folding \cite{Chan00}, but this is not our concern. We use the search for its ground 
states as a paradigmatic example for combinatorial optimization, with
a large body of existing benchmarks.

A wide variety of computational strategies have been employed to simulate and
analyze these models, including conventional (Metropolis) Monte Carlo
schemes with various types of moves \cite{yue95,rrp97,deutsch97}, chain growth
algorithms without \cite{tp92} and with re-sampling \cite{g98,bast98,zhang02}
(see also \cite{bd96}),
genetic algorithms \cite{um93,kd01}, parallel tempering \cite{irbaeck} and
generalizations thereof \cite{chikenji,chikenji2}, an `evolutionary Monte
Carlo' algorithm \cite{lw01}, and others \cite{tt96}. In addition, Yue and
Dill \cite{yd93_95} also devised an exact branch-and-bound algorithm
specific for HP sequences on cubic lattices, which gives all low
energy states
by exact enumeration and typically works for $N{<\atop\sim} 70 - 80$.

It is the purpose of the present letter to present a new variant of the
Pruned-Enriched Rosenbluth Method (PERM) \cite{g97} and to apply it to
lattice proteins. PERM is a biased chain growth algorithm with re-sampling
(``population control") and depth-first implementation.
It is built on the old idea of Rosenbluth and Rosenbluth (RR) \cite{rr55}
to use a biased growth algorithm for polymers, where the bias is corrected
by means of giving a weight to each sample configuration. While the chain
grows by adding monomers, this weight (which also includes the Boltzmann
weight if the system is thermal) will fluctuate.
PERM suppresses
these fluctuations by ``pruning" configurations with too low weight, and
by ``enriching" the sample with copies of high-weight configurations
\cite{g97}.
These copies are made while the chain is growing, and continue
to grow independently of each other. PERM can
be viewed as a special realization of a ``go with the winners" strategy
\cite{aldous} and indeed dates back to the beginning of the Monte Carlo
simulation era, when it was called ``Russian roulette and splitting"
\cite{kahn}. Among statisticians, this approach is also known as
sequential importance sampling (SIS) with re-sampling \cite{liu}.

Pruning and enrichment are done by choosing thresholds
$W_n^<$ and $W_n^>$ depending on the estimate of the partition sums
of $n$-monomer chains (see below for their actual determination).
If the current weight $W_n$ of an $n$-monomer chain is less than $W_n^<$,
the chain is discarded with probability 1/2, otherwise it is kept and its
weight is doubled. Many alternatives to this simple
choice are discussed in \cite{liu}, but we found that more sophisticated
strategies had little influence on the efficiency, and thus we kept the
above in the present work.
On the contrary, we found that different strategies in biasing and, most
of all, in enrichment had a big effect, and it is here the
present variant differs from those in \cite{g98,bast98}.
There, high-weight configurations were simply cloned
and the weight was uniformly shared between the clones.
For relatively high temperatures this is very efficient \cite{g97},
since each clone has so many possibilities to continue that different
clones very quickly become independent from each other. This is no
longer the case for very low temperatures. There we found that clones
often evolved in the same direction, since one continuation has a much higher
Boltzmann weight than all others. Thus, cloning is no longer efficient in
creating configurational diversity, which was the main reason why it was
introduced.

The main modification made in the present paper is thus that we no longer
make {\it identical clones}. Rather, when we have a configuration with $n-1$
monomers, we first estimate a {\it predicted} weight $W_n^{\rm pred}$ for
the next step, and we count the number $k_{\rm free}$ of free sites where
the $n$-th monomer can be placed. If $k_{\rm free}>1$ and
$W_n^{\rm pred}> W_n^>$,
we choose $2 \le k\le k_{\rm free}$ {\it different} sites among the free
ones and continue with $k$ configurations which are {\it forced} to be
different. Thus we avoid the loss of diversity which limited the success of
old PERM. Typically, we used $k = \min \{k_{\rm free},
\lceil W_n^{\rm pred}/W_n^>\rceil\}$.

When selecting a $k$-tuple $A = \{\alpha_1,\ldots \alpha_k\}$ of mutually
different continuations $\alpha_j$ with probability $p_A$,
the corresponding weights $W_{n,\alpha_1}\ldots, W_{n,\alpha_k}$ are
\be
   W_{n,\alpha_j} = \frac{ W_{n-1} q_{\alpha_j} k_{\rm free}}
                     { k {k_{\rm free}\choose k} p_A}\;.
\label{nis0}
\ee
where the {\it importance} $ q_{\alpha_j} = \exp(-\beta E_{n,\alpha_j}) $
of choice $\alpha_j$ is the Boltzmann-Gibbs factor associated with the
energy $E_{n,\alpha_j}$ of the newly placed monomer in the potential
created by all previous monomers. The other terms arise from correcting
bias and normalization, see \cite{hmng} for a more thorough discussion.
Choosing
\be
   p_A = {\sum_{\alpha \in A} q_\alpha \over \sum_{A'}\sum_{\alpha'\in A'}
         q_{\alpha'}}\;.                                     \label{nis}
\ee
would result in usual importance sampling \cite{hmng}. However,
instead of $q_{\alpha}$ we use the modified importances
$\tq_\alpha = (k_{\rm free}^{(\alpha)} + 1/2)q_{\alpha}$
in Eq.~(\ref{nis}), $k_{\rm free}^{(\alpha)}$ being the number of free
neighbors when the $n$-th monomer is placed at $\alpha$.
This replacement is made since we anticipate that continuations with less
free neighbours will contribute less on the long run than continuations
with more free neighbours.  This is similar to ``Markovian anticipation"
\cite{cylinder} within the framework of old PERM, where a bias different
from the short-sighted optimal importance sampling was found to be
preferable. Consequently, the predicted weight is
$W_n^{\rm pred} = W_{n-1}\sum_\alpha \tq_\alpha$,

A noteworthy feature of new PERM is that it crosses over to complete
enumeration when $W_n^<$ and $W_n^>$ tend to zero. In this limit, all
possible branches are followed and none is pruned as long as its weight is
not strictly zero. In contrast to this, old PERM would have made infinitely
many copies of the same configuration. This suggests already that we can be
more lenient in choosing $W_n^<$ and $W_n^>$. For the first configuration
hitting
length $n$ we used $W_n^<=0$ and $W_n^>=\infty$, i.e. we neither pruned nor
branched. For the following configurations we used $W_n^>=Z_n/Z_0 (c_n/c_0)^2$
and $W_n^<=0.2\,W_n^>$. Here, $c_n$ is the total number of configurations
of length $n$ already created during the run, and $Z_n$ is the partition sum
estimated from these configurations.

In PERM we work at a fixed temperature (no annealing), and successive
``tours" \cite{g97} are independent except for the thresholds
$W_n^{<,>}$ which use partially the same partition sum estimates.
Results are less sensitive to the precise choice of temperature
than they were for old PERM.  In general all temperatures in the range
$0.25 < T < 0.35$ gave good results for ground state search.
In the following, when we quote numbers of ground state hits
or CPU times between such hits, these are always {\it independent} hits.
The actual numbers of (dependent) hits are much larger.

We now present our results. Special comparison is made with the
{\it Core-directed Growth Method } (CG) of Beutler and Dill \cite{bd96},
the only method we found to be still competitive with ours. We emphasize,
however, that the CG method works only for the HP model and relies heavily on 
heuristics, in contrast to our fully blind general purpose approach.

{\bf (a)} We first tested the ten 48-mers from \cite{yue95}. As with old PERM,
we could reach lowest energy states for all of them, but within much shorter
CPU times. For
all 10 chains we used the same temperature, $\exp(1/T) = 18$, although we
could have optimized CPU times by using different temperatures for each
chain.

The CPU times for new PERM in Table~\ref{table1} are typically one order of
magnitude smaller than those in \cite{bd96}, except for sequence \#9 whose
lowest energy was not hit in \cite{bd96}. Since in \cite{bd96} a SPARC 1
machine was used which is slower by a factor $\approx 10$ than the 167 MHz
Sun ULTRA I used here, this means that our algorithms have comparable speeds.
We note that introducing a simple configurational bias in new PERM
\cite{footnote} can already give a considerable speedup;
in this contribution, however,
we want to concentrate on blind search.

\begin{table}
\begin{ruledtabular}
\caption{ Performances for the 3-d binary (HP-) sequences from
\protect\cite{yue95}.
} \label{table1}
\begin{tabular}{ccrrr}
 sequence &
 $-E_{\rm min}$\footnote{Ground state energies \protect\cite{yue95}.} &
 PERM\footnote{CPU times (minutes) per independent ground state hit,
      on 167 MHz Sun ULTRA I work station; from Ref.~\cite{bast98}} &
 new PERM\footnote{CPU times (minutes), same machine} &
 new PERM\footnote{CPU times (minutes), same machine} \\
   nr.    &          &         &         & with bias \cite{footnote} \\
   \hline
    1     &    32    &    6.9  &    0.63 &   0.13 \\
    2     &    34    &   40.5  &    3.89 &   0.23 \\
    3     &    34    &  100.2  &    1.99 &   0.71\\
    4     &    33    &  284.0  &   13.45 &   6.57 \\
    5     &    32    &   74.7  &    5.08 &   2.55 \\
    6     &    32    &   59.2  &    6.60 &   1.44 \\
    7     &    32    &  144.7  &    5.37 &   3.35 \\
    8     &    31    &   26.6  &    2.17 &   0.46 \\
    9     &    34    & 1420.0  &   41.41 &   10.53 \\
   10$\;\;$ &  33    &   18.3  &    0.47 &   0.08 \\
\end{tabular}
\end{ruledtabular}
\end{table}

{\bf (b)} Next we studied the two 2-d HP-sequences of length $N=100$ of
Ref.~\cite{rrp97}. They were originally thought to have ground states fitting
into a $10\times 10$ square with energies -44 and -46 \cite{rrp97}, but in
\cite{bast98} configurations fitting into this square were found with lower
energies. Moreover, when configurations were allowed to have arbitrary
shape, even lower energies were found
\cite{bast98,chikenji,zhang02}. In the present work we studied only
configurations of the latter type. The lowest energies known by now
are -48 \cite{zhang02} resp. -50 \cite{chikenji}. The CPU times needed
to find them were 48 min resp. 50h, on machines with $\approx 500$ MHz.
In contrast, new PERM needed in average 2.6 min resp. 5.8h on a
667 MHz DEC Alpha 21264 between any two hits.

{\bf (c)} Several 2-d HP-sequences were introduced in \cite{um93}, where the
authors tried to fold them using a genetic algorithm. Except for the shortest
chains they were not successful, but putative ground states for all of them
were found in \cite{bast98,irbaeck,chikenji}. But for the longest of these
chains ($N=64$), the ground state energy $E_{\rm min}=-42$ was
found in \cite{bast98} only by means
of special tricks which amount to non-blind search. With blind search, the
lowest energy reached by PERM was -39. We should stress that PERM as used in
\cite{bast98} was blind for all cases except this 64-mer (and when it found
$E=-49$ for the second $N=100$ chain of \cite{rrp97}), in contrast to
statements to the contrary made in \cite{lw01}.

We now found putative ground states for all chains of \cite{um93} with blind
search. For the 64-mer the average CPU time per hit was ca. 30h on the
DEC 21264, which seems to be roughly comparable to the CPU times
needed in \cite{chikenji,irbaeck}, but considerably slower than \cite{bd96}.
This sequence is particularly difficult for any growth
algorithm, and the fact that we now found it is particularly
noteworthy.

On the other hand, new PERM was much faster than \cite{bd96} for the sequence
with $N=60$ of \cite{um93}. It needed $\approx 10$ seconds on the DEC 21264
to hit $E_{\rm min}=-36$, and $\approx 0.1$ second to hit $E=-35$. In contrast,
$E=-36$ was never hit in \cite{bd96}, while it took 97 minutes to hit $E=-35$.

{\bf (d)} A 85-mer 2-d HP sequence was given in \cite{konig}, where it was
claimed to have $E_{\rm min}=-52$. Using a genetic algorithm, the authors
could find only conformations with $E\ge -47$. In Ref.~\cite{lw01}, using
a newly developed {\it evolutionary Monte Carlo} (EMC) method, the authors
found the putative ground state when assuming large parts of its known
structure as constraints. This amounts of course to non-blind search.
Without these constraints, the putative ground state was not hit in
\cite{lw01} either, although the authors claimed their algorithm to be more
efficient than all previous ones. We easily found states with $E=-52$, but
we also found many conformations with $E=-53$. At $\exp(1/T)=90$
it took ca. 10 min CPU time between successive hits on the Sun ULTRA 1.

{\bf (e)} Four 3D HP sequences with $N=58$, $103$, $124$, and $136$ were
proposed in \cite{dfc93,lfd94} as models for actual proteins or protein
fragments. Low energy states for these sequences were searched
in \cite{tt96} using a newly developed and supposedly very
efficient algorithm. The energies reached in \cite{tt96} were $E=-42$,
$-49$, $-58$ and $-65$, respectively. We now found lower
energy states after only few minutes CPU time, for all four chains. For
the longer ones, the true ground state energies are indeed {\it much} lower
than those found in \cite{tt96}, see Table~\ref{table2}.

Note the very low temperatures needed to fold the very longest chains in
an optimal time. If we would be interested in excited states, higher
temperatures would be better. For instance, to find $E=-66$ for the
136-mer (which is one unit below the lowest energy reached in \cite{tt96}),
it took just 2.7 seconds/hit on the DEC 21264 when using $\exp(1/T)=40$.

\begin{table*}
\caption {\label{table2} Newly found lowest energy states for binary
      sequences with interactions
$\vec{\epsilon}=(\epsilon_{HH},\epsilon_{HP},\epsilon_{PP})=-(1,0,0)$.}
\begin{ruledtabular}
\begin{tabular}{rrcllrl}
  &     &          & old & new & & CPU\\
N & $d$ & Sequence & $E_{\rm min}{\rm [Ref.]}$ &$E_{\rm min}$ &
  $ {\displaystyle e^{1\over T}} $ &
  time\footnote{hours per independent hit on 667 MHz DEC ALPHA 21264} \\
  & & {\it example conformation\footnote{r=right, l=left, f=forward, b=backward, u=up, d=down}} & & & & \\
\hline
85 & 2 &
$H_4P_4H_{12}P_6H_{12}P_3H_{12}P_3H_{12}P_3HP_2H_2P_2H_2P_2HPH$
& -52\cite{lw01} & -53  & 90 & 0.03 \\
& & $flb_3lf_4lf_2rbrbrfr_2f_3l_2b_2lf_2lbl_2frfl_2b_2lbr_2b_3rb_3l_2frflf_3lb_5lf_2lfrflfrflfrfr$ & & & & \\
58 & 3 &
$PHPH_3PH_3P_2H_2PHPH_2PH_3PHPHPH_2P_2H_3P_2HPHP_4HP_2HP_2H_2P_2HP_2H$
& -42\cite{tt96} & -44 & 30 & 0.19 \\
 &  &
$ublfl_2urfldrfrbrub_2lf_3lublbrurdfrubdblbufldblfldr_2bdfdlu$
& & & & \\
103 & 3 &
$P_2H_2P_5H_2P_2H_2PHP_2HP_7HP_3H_2PH_2P_6HP_2HPHP_2$
& -49\cite{tt96} &  -54\cite{footnote} & 60 & 3.12 \\
  &    & $HP_5H_3P_4H_2PH_2P_5H_2P_4H_4PHP_8H_5P_2HP_2$ & & & & \\
&  & 
$ufrbdflfurdfu_2rd_2buruf_2ulbluld_2burdrubrdl_2bufldblfulf_2rd$ & & & &\\
& & $ bd_2b_2uflufd_3fu
rurd_2fu_2ru_2ldf_2urbl_2dbdlbulfru_2$ 
& & & & \\
124 & 3 &
$P_3H_3PHP_4HP_5H_2P_4H_2P_2H_2P_4HP_4HP_2HP_2H_2P_3H_2PHPH_3P_4H_3P_6$
& -58\rm \cite{tt96} &  -71 & 90 & 12.3$\;\;$ \\
  &    & $H_2P_2HP_2HPHP_2HP_7HP_2H_3P_4HP_3H_5P_4H_2PHPHPHPH$ & & & & \\
& & 
$urbd_2bublfurb_2drf_5ub_4ufluldfrufrbdfrbubd_2burf_3dlbrb_3df_2$ & & & &\\
& & $lf_3urdb_2d_2luflb_2rbrfdrfrubulbuf_2u_2b_2dfrbdf_2dldf_2u_2bdrurbulfl$ 
& & & & \\
136 & 3 &
$HP_5HP_4HPH_2PH_2P_4HPH_3P_4HPHPH_4P_{11}HP_2HP_3HPH_2P_3H_2P_2HP_2$
& -65\cite{tt96} &  -80 & 120 & 110$\;\;\;\;\;$ \\
  &    &  $HPHPHP_8HP_3H_6P_3H_2P_2H_3P_3H_2PH_5P_9HP_4HPHP_4$   & & & & \\
& &
$u_2b_2rdl_2frbdrdlf_2lfr_2brblu_3fd_2rbubd_2r_2df_3dl_2ul_2blbrdfrurbldrbul_2b$ & & & & \\
& & $rdrurf_2urf_2ububdlu_2bd_2blurul_2d_2ldr_4ubld_2l_2urubu_3brd_2f_2u_2ld_2ldbubu$ & & & & \\
\end{tabular}
\end{ruledtabular}
\end{table*}

{\bf (f)} The only case where we could not find a known ground state is a 3-d 
HP sequence of length 88 given in
\cite{bd96}. As shown there, it folds into an irregular $\beta/\alpha$-barrel
with $E_{\rm min}=-72$. The
difficulties of PERM with this sequence are easily understood by looking at
the configuration shown in \cite{bd96}. The nucleus of the hydrophobic
core is formed by amino acids \#36-53. Before its formation, a growth algorithm
starting at either end has to form very unstable and seemingly unnatural
structures which are stabilized only by this nucleus,
a situation similar to the 64-mer of Ref.~\cite{um93}.
In order to fold also
this chain, we would have either to start from the middle of the chain (as
done in \cite{bast98} for some sequences) or use some other heuristics which
help the formation of the hydrophobic core. Since we wanted our algorithm to be
as general and ``blind" as possible, we did not incorporate such tricks
\cite{footnote}.

A more detailed discussion of our algorithm, the results,
and comparison with other methods  is given elsewhere \cite{hmng}.
A list containing all sequences for which we found new lowest energy
configurations is given in Table~\ref{table2}.

In the present paper we presented a new version of PERM which is a
depth-first
implementation of the `go-with-the-winners' strategy (or sequential importance
sampling with re-sampling). The main improvement over old PERM
is that we now do not make
{\it identical clones} of high weight (partial) configurations, but we branch
such that each continuation is forced to be different. We do not expect this to
have much influence for systems at high temperatures, but as we showed, it
leads to substantial improvement at very low temperatures.

Comparing our results to previous work, we see that we found the known lowest
energy states in {\it all} cases but one. Moreover, whenever we could compare
with previous CPU times, the comparison was favourable for our new algorithms,
except for the CG method of Beutler and Dill \cite{bd96}. But we should
stress that the latter is very specific to HP chains, uses strong heuristics
regarding the formation of a hydrophobic core, and does not give correct
Boltzmann weights for excited states. All that is not true for our method.

Although our method could be used for a much wider range of applications (see
\cite{chemnitz} for applications of PERM), we presented here only results for 
heteropolymers with two types of monomers and the simplest non-trivial 
interactions on the square and simple cubic lattices. But we applied it also 
successfully to the HP model on the FCC lattice, to off-lattice heteropolymers,
and to lattice models with more than 2 types of monomers (to be published).
We hope that our results will also foster applications to more realistic
protein models. We showed only results for lowest energy configurations, but we
should stress that PERM is not only an optimization algorithm.
It also gives information on the full thermodynamic behaviour. We skipped this
here since finding ground states is the most difficult problem in general, and
sampling excited states is easy compared to it.


\begin{thebibliography}{30}
\bibitem{footnote} In additional runs
   we anticipated that contacts between hydrophobic and polar residues should
   be rare in native states. We thus punished each such contact by
   multiplying the weight with a factor $q_{HP} < 1$. Optimal values of
   $q_{HP}$ depended on the sequence, but $q_{HP} = 0.2$ never was detrimental 
   and typically reduced the CPU times by factors between 2 and 10. For the
   103-mer from \cite{dfc93,lfd94,tt96} we even found a new lowest energy,
   $E=-55$.
\end{thebibliography}
\end{document}